\documentclass[showpacs, onecolumn,superscriptaddress]{revtex4}

\usepackage{doi}
\usepackage{hyperref}
\hypersetup{
  colorlinks=true,        
  linkcolor=blue,         
  citecolor=cyan,         
}

\usepackage{graphicx}
\usepackage{dcolumn}
\usepackage{bm}
\usepackage{color}
\usepackage{enumitem}
\usepackage{amsmath}
\usepackage{amssymb}
\usepackage{orcidlink}
\usepackage{soul}
\usepackage{xcolor}

\begin{document}

\title{Repetitive Penrose Process in Kerr-Taub-NUT black hole spacetime}

\author{Mirzabek Alloqulov,\orcidlink{0000-0001-5337-7117}}
\email{malloqulov@gmail.com}
\affiliation{School of Physics, Harbin Institute of Technology, Harbin 150001, People’s Republic of China}

\affiliation{University of Tashkent for Applied Sciences, Str. Gavhar 1, Tashkent 100149, Uzbekistan}

\affiliation{New Uzbekistan University, Movarounnahr str. 1, Tashkent 100000, Uzbekistan}

\author{Bobomurat Ahmedov,\orcidlink{0000-0002-1232-610X}}
\email{ahmedov@astrin.uz}

\affiliation{School of Physics, Harbin Institute of Technology, Harbin 150001, People’s Republic of China}
\affiliation{Institute for Advanced Studies, New Uzbekistan University, Movarounnahr str. 1, Tashkent 100000, Uzbekistan}

\affiliation{Institute of Theoretical Physics, National University of Uzbekistan, Tashkent 100174, Uzbekistan}

\author{Chengxun Yuan,\orcidlink{0000-0002-2308-6703}}
\email{yuancx@hit.edu.cn}

\affiliation{School of Physics, Harbin Institute of Technology, Harbin 150001, People’s Republic of China}

%
\date{\today}
\begin{abstract}

In this article, we study the repetitive Penrose process for the Kerr-Taub-NUT black hole (BH). First of all, we briefly review the spacetime of the Kerr-Taub-NUT BH, including horizon and ergosphere structures. The results indicate that the event horizon and ergosphere radii increase under the influence of the gravitomagnetic charge $l$. Subsequently, we find by using the irreducible mass of the BH that the extractable energy decreases with the rise of the gravitomagnetic charge. 
We then turn to the repetitive Penrose process by writing the conservation laws and setting the corresponding iterative stopping conditions. Furthermore, we numerically calculate the change in the BH's parameters, along with the corresponding quantities of the repetitive Penrose process, for each iteration. 
\end{abstract}

\maketitle

\section{Introduction}

The extraction of energy from spinning black holes (BHs) remains one of the most intriguing predictions of general relativity (GR) and a cornerstone in the investigation of high-energy astrophysical phenomena. Among the various mechanisms, the Penrose process~\cite{Penrose,Penrose:1969pc} stands out as a fundamental theoretical framework that demonstrates how rotational energy can be extracted from a BH via particle interactions within the ergoregion. In other words, the parent particle enters the ergoregion of the BH and splits into two fragments, one of which falls to the BH horizon, and the other goes to infinity with an energy higher than the energy of the parent particle. Nevertheless, for this process to take place, the relative three-dimensional speed of the splitted particles must be greater than half the speed of light, a condition that is difficult to attain in reality~\cite{Bardeen:1972fi,Wald:1974kya}. In realistic astrophysical environments, BHs are usually surrounded by plasma and strong electromagnetic fields, which give rise to more efficient and physically relevant mechanisms. The electric, magnetic, and collisional Penrose processes~\cite{PhysRevD.29.2712,PhysRevD.30.1625,1985ApJ...290...12W,Bhat:1985hpc,Wagh:1989zqa,PhysRevLett.109.121101,Dadhich_2018,shaymatov2024energeticsbuchdahlstarsmagnetic,1986ApJ...307...38P,Xamidov:2024wou,Shaymatov:2024fle,Schnittman_2018,Viththani:2024map,Patel:2023efv,Zhang:2025blr,Tursunov:2019oiq,Shaymatov22b,Tursunov20ApJ,PhysRevD.104.084099,Alloqulov2023302,Xamidov:2025pln,Alloqulov:2024cto,Kurbonov:2023uyr} extend the original idea by incorporating the interaction of charged particles with electromagnetic fields or utilizing high-energy particle collisions, significantly enhancing the achievable energy gain. In parallel, the Blandford-Znajek mechanism~\cite{1977MNRAS.179..433B} offers a fundamentally different, yet complementary, framework in which rotational energy is extracted electromagnetically through magnetic field lines threading the event horizon. This process is widely regarded as one of the primary engines powering relativistic jets in accreting BH systems. Moreover, the magnetic reconnection~\cite{Koide:2008xr,Comisso21,Yao:2026cvs,Shaymatov2024,Zeng:2025olq,Long:2024tws,Rodriguez:2024jzw,Zhang:2024rvk,Ye:2023xyv,Li:2023nmy,Wang:2022qmg,Carleo:2022qlv,Cheng:2025qlc} in highly magnetized plasma environments provides an efficient channel for converting magnetic energy into particle energy, often operating in tandem with these extraction mechanisms and further amplifying energetic outputs. 

Recently, Ruffini et al.~\cite{Ruffini:2024dwq} derived turning point constraints on particle trajectories in the Kerr spacetime within the framework of the original Penrose process, showing that energy extraction remains feasible under corresponding conditions. Subsequently, they introduced a repetitive version of the Penrose process~\cite{Ruffini:2024irc}. Their findings revealed that the repetitive process does not permit the extraction of the entire rotational energy reservoir of the Kerr BH; instead, part of the extractable energy is transformed into irreducible mass. This bound originates because after each extraction cycle, the updated mass and spin of the BH determine the available energy for subsequent extractions, with the irreducible mass nonlinearly increasing. Subsequently, this framework was extended to the repetitive electro-Penrose process~\cite{Hu_2026}, and to other BH spacetimes~\cite{Zeng:2026jpn,Zeng:2026egq,Wang:2025nfv,Hu:2025bbc}. 
        
While the repetitive Penrose process has been thoroughly explored in the context of the Kerr metric, a significant gap remains in our understanding of how it is modified in the presence of a gravitomagnetic monopole. The Kerr-Taub-NUT (Newman-Unti-Tamburino) spacetime, which generalizes the Kerr solution by including the NUT parameter $l$, describes a BH endowed not only with the mass and spin but also with a gravitomagnetic charge~\cite{Taub:1950ez,1966BAPSS..14..653D,Carter:1966zza,Bonnor_1969,Dadhich_2002}. This parameter introduces a non-trivial topological structure and significantly alters the geometry of the ergosphere as well as the horizon structure. The Kerr-Taub-NUT spacetime has been employed to investigate the characteristics of the axially symmetric BHs and holds significance for a range of astrophysical processes, including optical properties, accretion disks, and etc~\cite{Pradhan:2021pkx,Ghasemi-Nodehi:2021ipd,Sen:2022hzo,Perry:2022udk,Zhou:2022eiv,Narzilloev:2023btg,Lee:2023jfi,Alibekov:2024cdc,Sen:2024dct,Siahaan:2024ljt,Siahaan:2024zhx,Zi:2025idv,Duztas:2025kdb,Wang:2025edt}.

The present work seeks to bridge this gap by investigating the repetitive Penrose process within the Kerr-Taub-NUT BH spacetime. Our investigation begins with the review of the spacetime of the Kerr-Taub-NUT BH, including the ergosphere and horizon structures. Subsequently, we construct the iterative stopping conditions and analyze how the presence of the NUT parameter modifies the corresponding quantities of the repetitive Penrose process.

The present paper is structured in the following way: we briefly review the spacetime of the Kerr-Taub-NUT BH and revisit the theoretical framework of the repetitive Penrose process in Section~\ref{sec:2}. Section~\ref{sec:conditions} is devoted to the corresponding iterative stopping conditions. In Section~\ref{sec:numerical}, we present our numerical results for the BH parameters and the corresponding quantities of the repetitive Penrose process. Section~\ref{summary} contains our conclusions and discussions.

\section{Spacetime and theoretical framework of repetitive Penrose process}\label{sec:2}

One can write the spacetime of the Kerr-Taub-NUT BH in the following form~\cite{Taub:1950ez,1966BAPSS..14..653D,Carter:1966zza,Bonnor_1969,Dadhich_2002}
\begin{equation}
ds^2 = -\frac{1}{\Sigma} (\Delta - a^2 \sin^2 \theta) dt^2 + \frac{\Sigma}{\Delta} dr^2 + \Sigma d\theta^2 \\
+ \frac{1}{\Sigma} [(\Sigma + a\chi)^2 \sin^2 \theta - \chi^2 \Delta] d\phi^2 \\
+ \frac{2}{\Sigma} [\Delta\chi - a(\Sigma + a\chi)\sin^2\theta] d\phi dt,
\end{equation}
with
\begin{align}
\Delta=r^2+a^2-l^2-2 M r\,, \qquad \Sigma=r^2+(l+a\cos{\theta})^2\,, \qquad \chi=a\sin^2{\theta}-2l\cos{\theta}\,.
\end{align}
where $M$ and $a$ refer to the BH's mass and spin, respectively, while $l$ represents the gravitomagnetic charge. Note that one can recover the Kerr BH when gravitomagnetic charge tends to zero, i.e., $l\rightarrow0$. 
By solving the $g^{rr}=0$, one can obtain the event horizon radii of the Kerr-Taub-NUT BH as follows
\begin{equation}
    r_{\pm}=M \pm \sqrt{M^2-a^2+l^2}\,.
\end{equation}
 In addition, we can get the expression for the outhor ergoregion surface by solving $g_{tt}=0$ as
\begin{equation}
r_{E}=M+\sqrt{M^2-a^2\cos^2{\theta}+l^2}\,.
\end{equation}
After that, one can find the surface area of the event horizon as follows~\cite{Wang:2025nfv}
\begin{equation}
S=\int_0^{\pi}\int_0^{2\pi} \sqrt{g_{\theta\theta}g_{\phi\phi}}d\phi d\theta=4 \pi (r_+^2+a^2)\,.
\end{equation}
Using the above equation, the irreducible mass can be found in the following form~\cite{Hu:2025bbc}
\begin{equation}
    M_{irr}=\sqrt{\frac{S}{16\pi}}=\frac{1}{2}\sqrt{{r_{+}^2+a^2}}\,.
\end{equation}
We then write the extractable energy as
\begin{equation}
    E_{extractable}=M-M_{irr}=M-\frac{1}{2}\sqrt{{r_{+}^2+a^2}}\,.
\end{equation}
The above expression represents the theoretically maximum extractable energy from the Kerr-Taub-NUT BH. 
\begin{figure}
    \centering
    \includegraphics[scale=0.32]{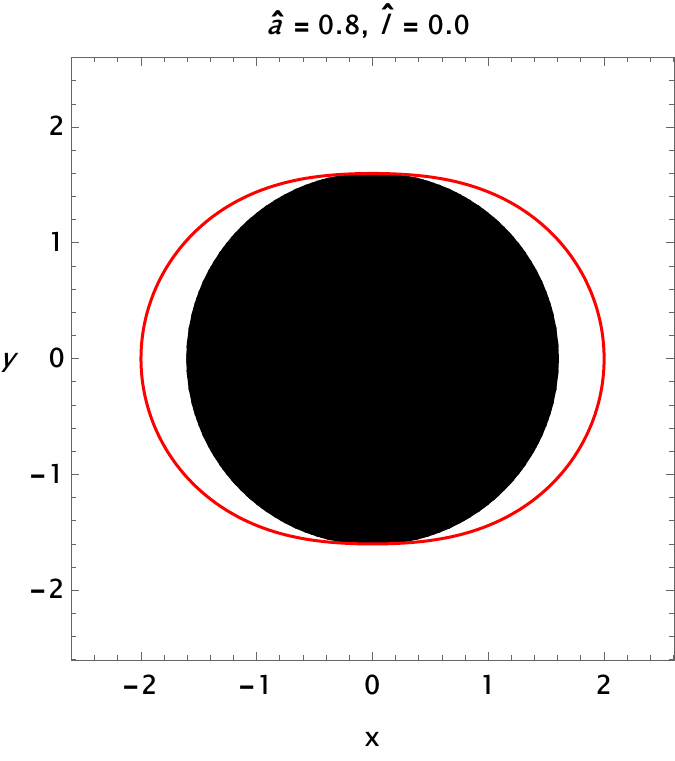}
    \includegraphics[scale=0.32]{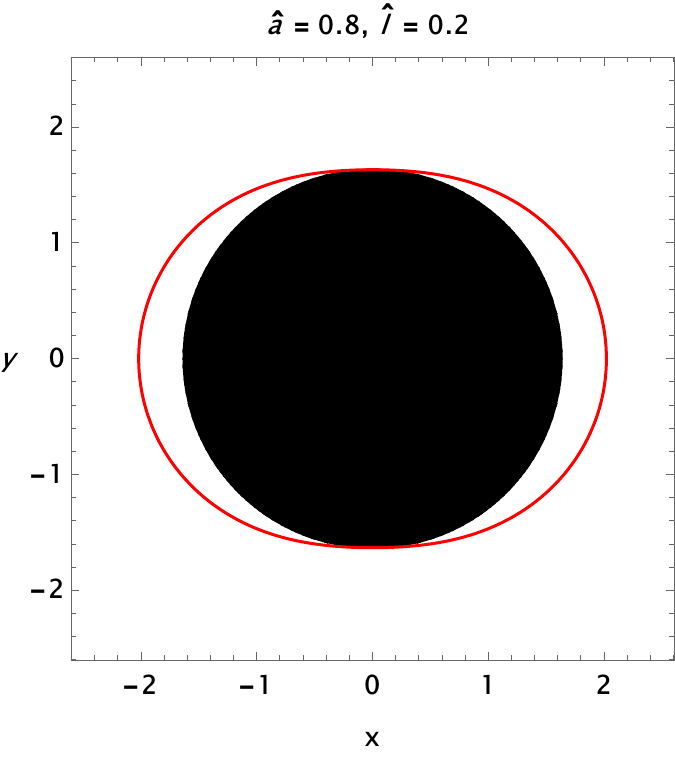}
    \includegraphics[scale=0.32]{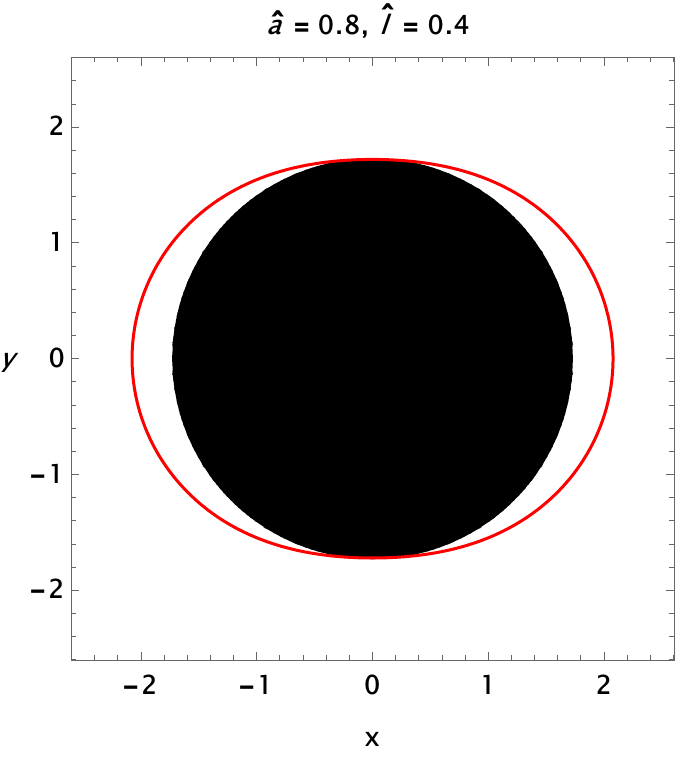}
    \includegraphics[scale=0.32]{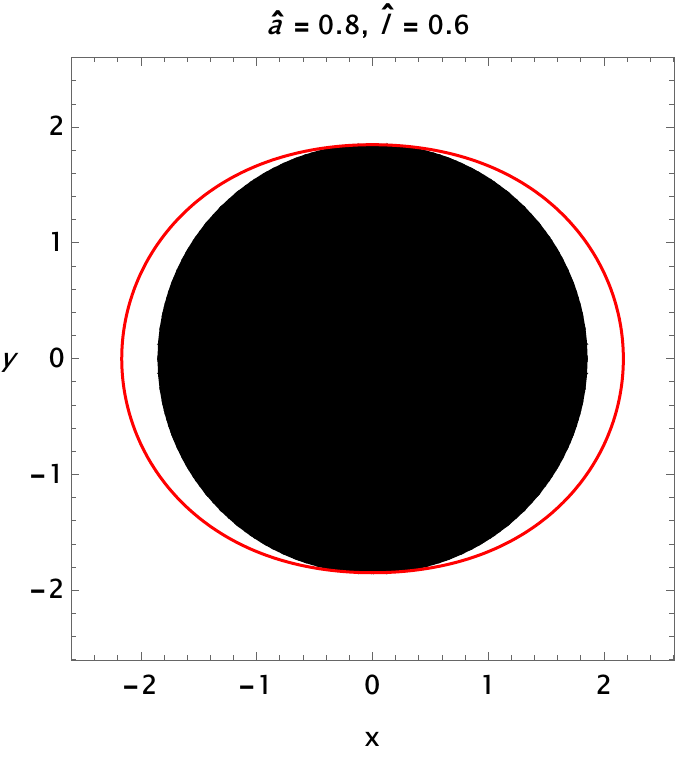}
    \includegraphics[scale=0.32]{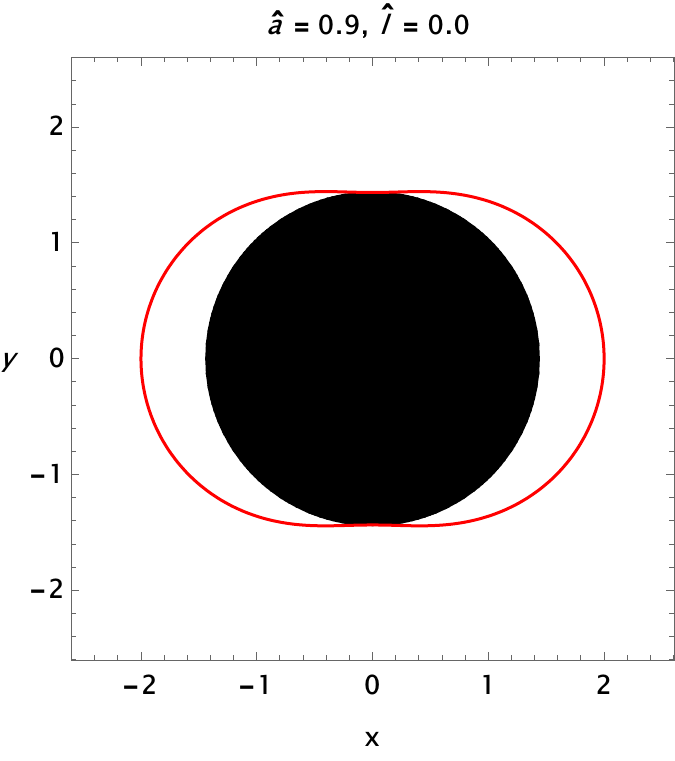}
    \includegraphics[scale=0.32]{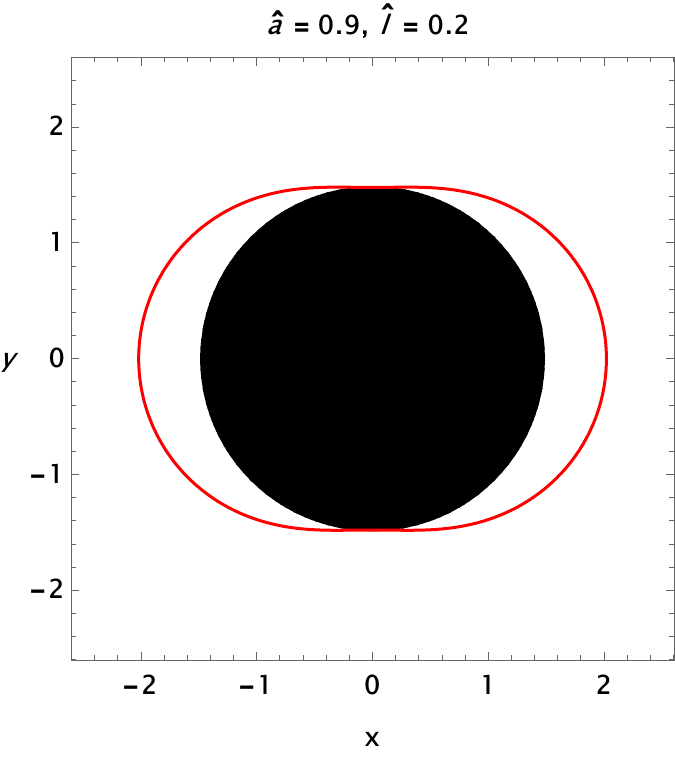}
    \includegraphics[scale=0.32]{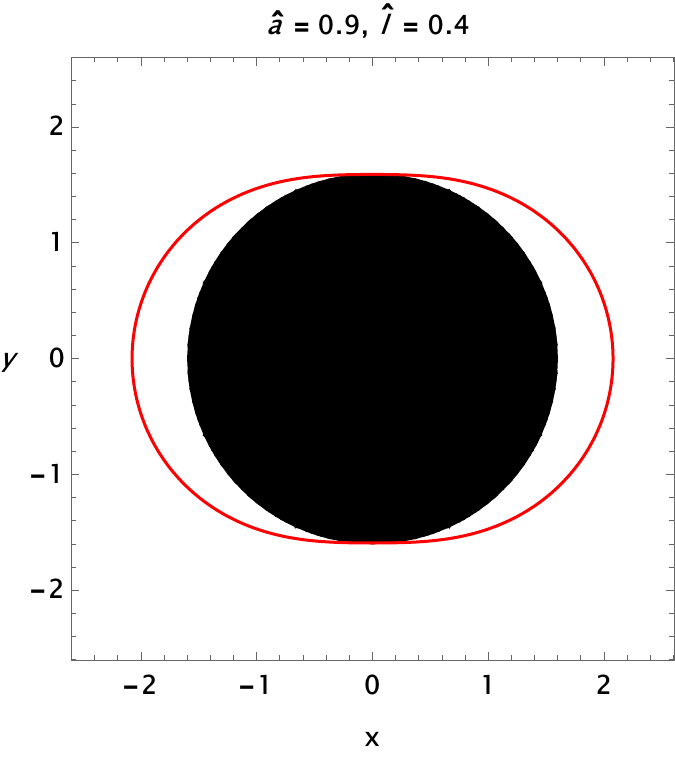}
    \includegraphics[scale=0.32]{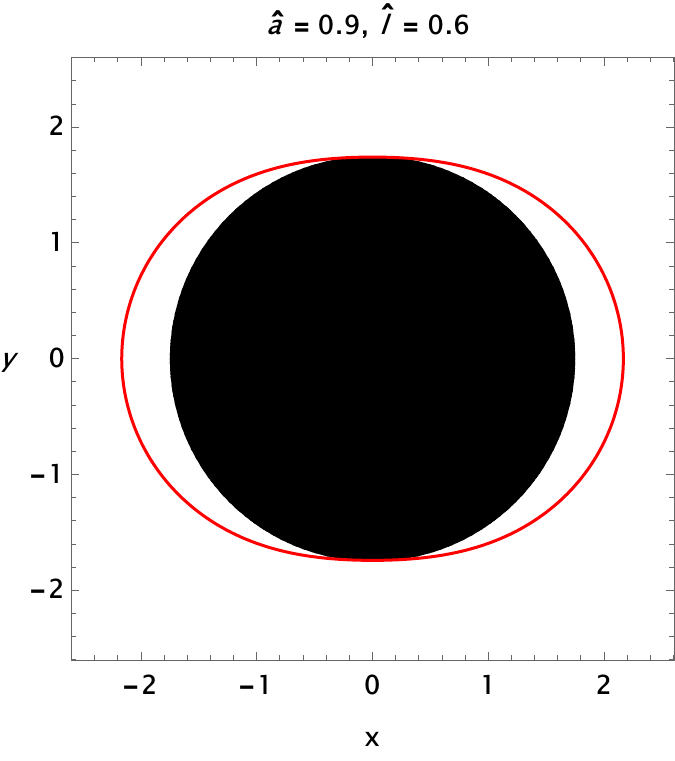}
    \includegraphics[scale=0.32]{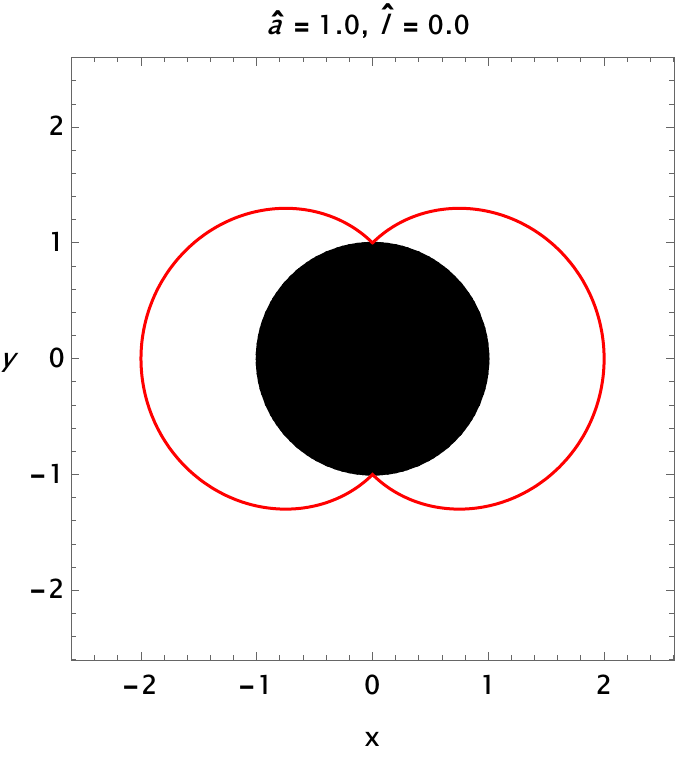}
    \includegraphics[scale=0.32]{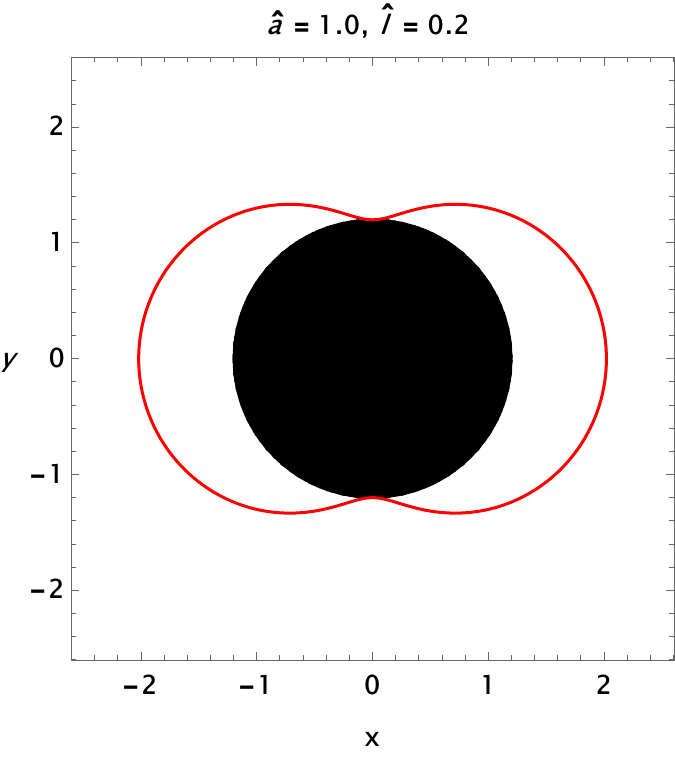}
    \includegraphics[scale=0.32]{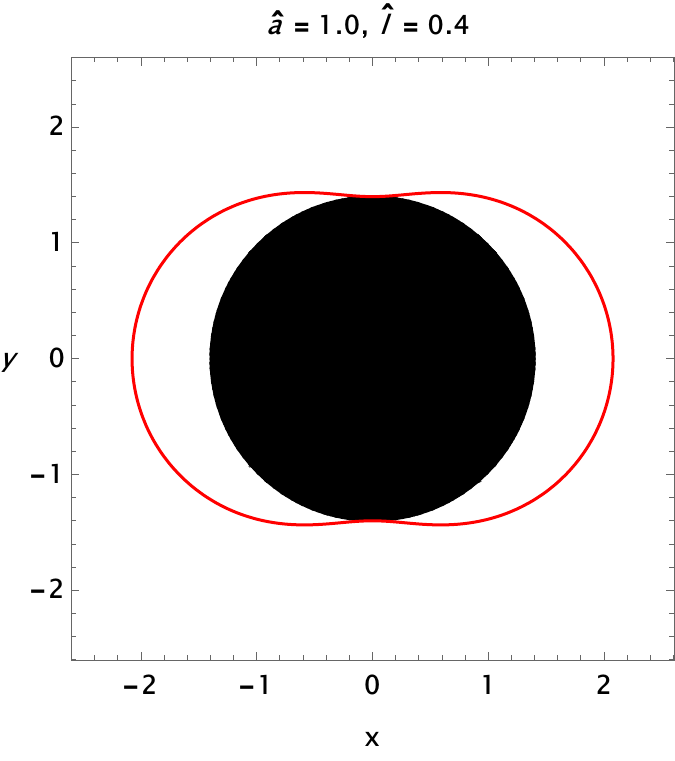}
    \includegraphics[scale=0.32]{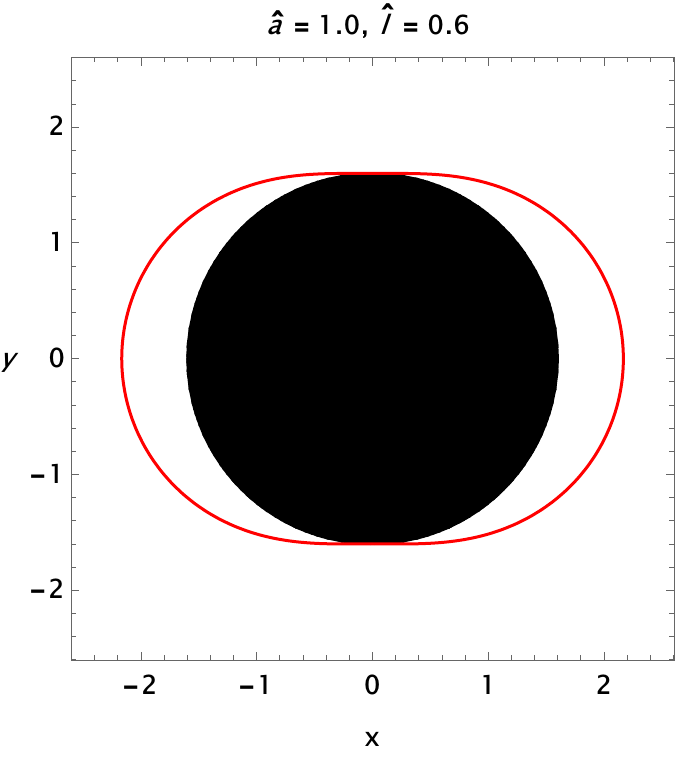}
    \caption{The plot shows the ergoregion of the Kerr-Taub-NUT BH for various combinations of the spin and gravitomagnetic charge of the BH.}
    \label{fig:ergo}
\end{figure}

\begin{figure}
    \centering
    \includegraphics[scale=0.5]{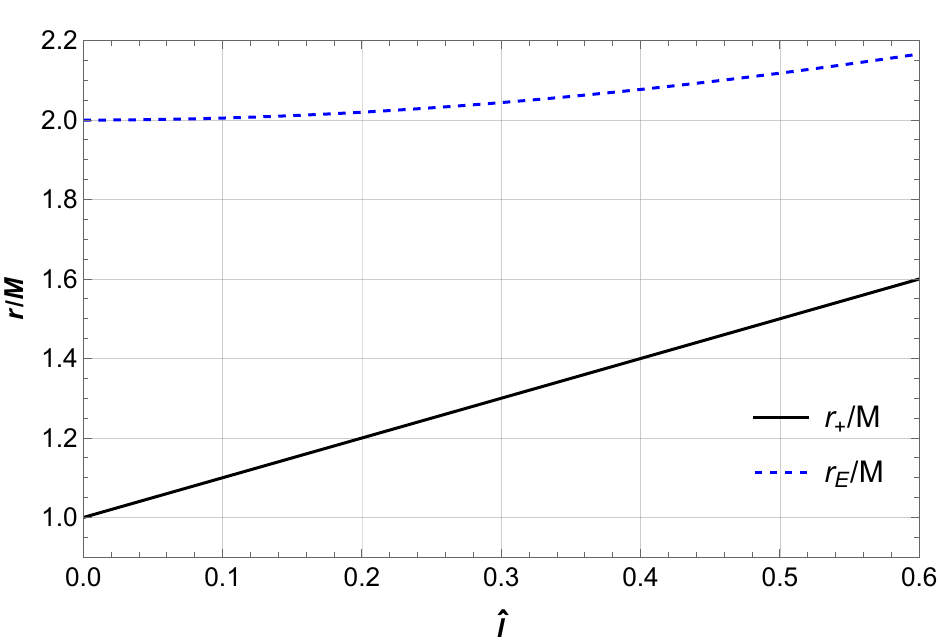}
    \includegraphics[scale=0.5]{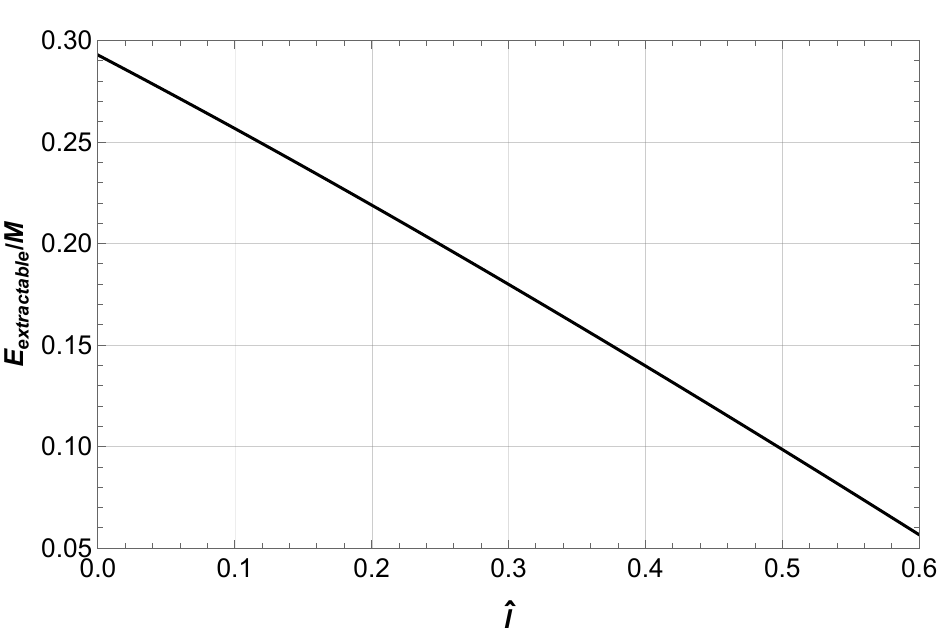}
    \caption{Left panel: the dependence of the event horizon and ergosphere on the dimensionless $\hat{l}$ parameter. Here, the spin of the BH is $a/M=1$. Right panel: the extractable energy of the Kerr-Taub-NUT BH as a function of the $\hat{l}$ parameter.}
    \label{fig:horizon}
\end{figure}
In Fig.~\ref{fig:ergo}, we demonstrate the ergoregion of the Kerr-Taub-NUT BH for the different values of the BH spin and the $l$ parameter. Subsequently, we plot the radius of the event horizon and outer ergoregion surface as a function of the $l$ parameter for the maximally spinning Kerr-Taub-NUT BH in the left panel of Fig.~\ref{fig:horizon}. It can be seen from this panel that there is an increase under the influence of the $l$ parameter. The right panel of Fig.~\ref{fig:horizon} illustrates the dependence of the extractable energy on the values of the gravitomagnetic charge. The values of the extractable energy decrease with the rise of the $l$ parameter.

Now, we can consider the Penrose process, and the conservation laws of energy, angular and radial momenta can be written as follows~\cite{Ruffini:2024dwq,Ruffini:2024irc}
\begin{align}
\hat{E}_0=\tilde{\mu}_1\hat{E}_1+\tilde{\mu}_2\hat{E}_2\,, \qquad \hat{p}_{\phi 0}=\tilde{\mu}_1 \hat{p}_{\phi 1}+\tilde{\mu}_2 \hat{p}_{\phi 2}\,, \qquad \hat{p}_{r 0}=\tilde{\mu}_1 \hat{p}_{r 1}+\tilde{\mu}_2 \hat{p}_{r 2}\,,
\end{align}
with
\begin{align}
\tilde{\mu}=\frac{\mu_i}{\mu_0},\,\,\, \hat{E}_i=\frac{E_i}{\mu_i},\,\,\, \hat{p}_{\phi i}=\frac{p_{\phi i}}{\mu_i M}, \,\,\, \hat{p}_{r i}=\frac{p_{r i}}{\mu_i}\,,
\end{align}
where $i$ runs from 0 to 2, while $\mu_i$ denotes to the mass of the $i$th particle. Using the equations of motion, one can find that the effective potential of the particle is moving radially in the equatorial plane ($\theta=\pi/2$) as~\cite{Wang:2025nfv}
\begin{equation}
\hat{V}_{eff i}^{\pm}=\frac{g^{t\phi}\hat{p}_{\phi i} \mp \sqrt{(g^{t\phi})^2\hat{p}^2_{\phi i}-g^{tt}(g^{\phi \phi}\hat{p}^2_{\phi i}+1)}}{g^{tt}}\,.
\end{equation}
As stated in Refs.~\cite{Zeng:2026jpn,Zeng:2026egq,Wang:2025nfv,Hu:2025bbc}, we consider $\hat{E}_i=\hat{V}_i^+$ case since all three particles need to be at their turning points, and the radial momentum of each of the three particles must vanish at the decay location. We further use the analytic expressions for $\hat{p}_{\phi 0}$, $\hat{p}_{\phi 2}$, $\hat{E}_1$ and $\hat{E}_2$ as obtained in~\cite{Wang:2025nfv}. 

As energy is extracted during each cycle of the repetitive Penrose process, the remaining mass and angular momentum of the Kerr-Taub-NUT black hole change as follows~\cite{Zeng:2026jpn,Zeng:2026egq}
\begin{align}
   M_n=M_{n-1}+\Delta M_{n-1}=M_{n-1}+\hat{E}_{1,n-1}\mu_{1,n-1}\,,  \qquad L_n=L_{n-1}+\hat{p}_{\phi 1}\mu_{1,n-1}M_{n-1}\,,
\end{align}
with $L_0=\hat{a}_0M_0^2$. After that, the BH's dimensionless spin $\hat{a}=a/M$ changes as
\begin{align}
    \Delta \hat{a}_{n-1}=\frac{L_n}{M^2_n}-\frac{L_{n-1}}{M^2_{n-1}}\,.
\end{align}
As an outcome, the irreducible mass also changes as
\begin{equation}
\Delta E_{extractable, n-1}=\Delta M_{n-1}-\Delta M_{irr,n-1}\,.
\end{equation}
Now, we can write the extracted energy through this process in the $n$ th cycle as $E_{extracted,n}=M_0-M_n$. It is worth noting that the effectiveness of energy extraction can be characterized by the quantities $\xi_n$ and $\Xi_n$, which denote the energy return on investment and energy utilization efficiency, respectively. They were defined through the following equations~\cite{Zeng:2026jpn,Zeng:2026egq}
\begin{align}
\xi_n=\frac{E_{extracted,n}}{nE_0}\,, \quad \Xi_n=\frac{E_{extracted,n}}{E_{extractable,0}-E_{extractable,n}}\,.
\end{align}

\section{Iterative stopping conditions}\label{sec:conditions}

The repetitive energy extraction process must fulfill the following conditions~\cite{Ruffini:2024dwq,Ruffini:2024irc}
\begin{itemize}
    \item The mass deficit must fulfill $1-\tilde{\mu}_1-\tilde{\mu}_2>0$ condition. 

    \item $\hat{E}_1$ must be lower than zero during the iteration process, i.e., $\hat{E}_1<0$. 

    \item $E_{extractable,n}$ must be higher than zero for each iteration, i.e., $E_{extractable,n}>0$. 

    \item The irreducible mass can not diminish across iterations. 

    \item The turning points of particles 0 and 2 must be located to the right of the maximum of the effective potential, and vice versa for particle 1. 
\end{itemize}

The last condition occurs when the classical turning point of each particle is located at the maximum of their effective potential, which is written as
\begin{align}
    \hat{V}_{eff,i}^+(\hat{r}_p)=\hat{E}_i\,, \quad \frac{{dV}_{eff,i}^+}{d\hat{r}}\Bigg|_{\hat{r}=\hat{r}_p}=0\,.
    \label{eff}
\end{align}
Here, $\hat{r}_p$ represents the dimensionless decay radius, i.e., $\hat{r}_p=r_p/M$. When $\hat{E}_0>1$, the condition for terminating the iteration at this stage is controlled by particle 0, which provides the minimum allowable spin. This threshold is given by the co-rotating marginally bound orbit of this particle. 

\begin{figure}
    \centering
    \includegraphics[scale=0.33]{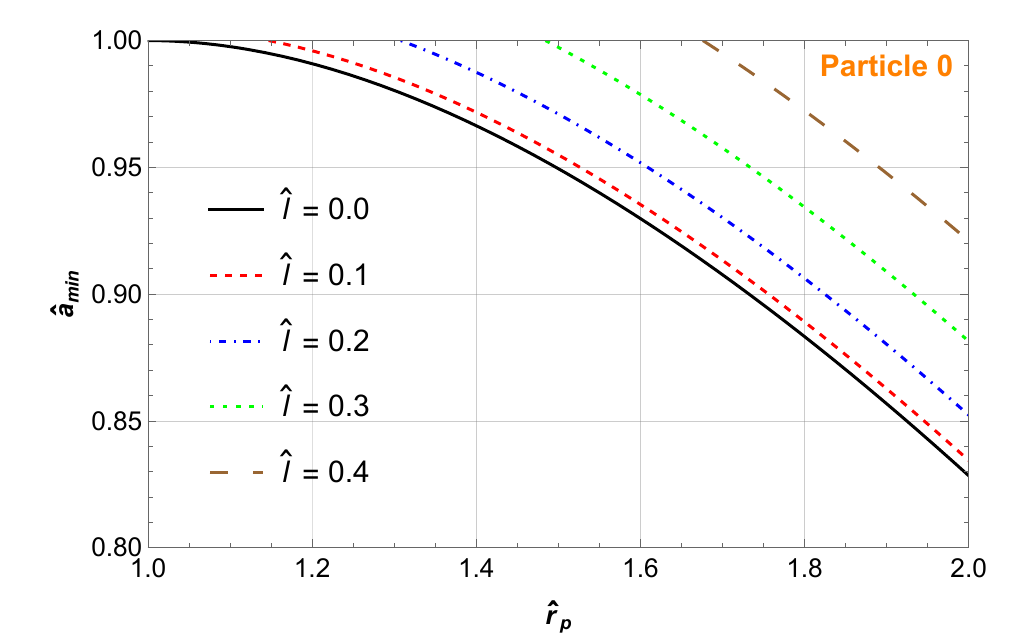}
    \includegraphics[scale=0.33]{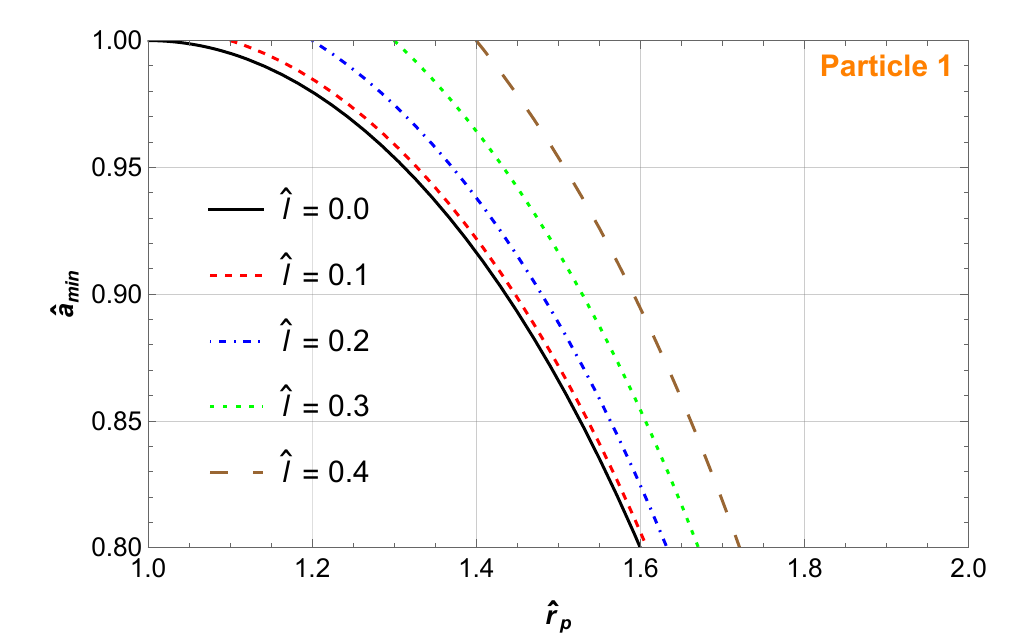}
    \includegraphics[scale=0.33]{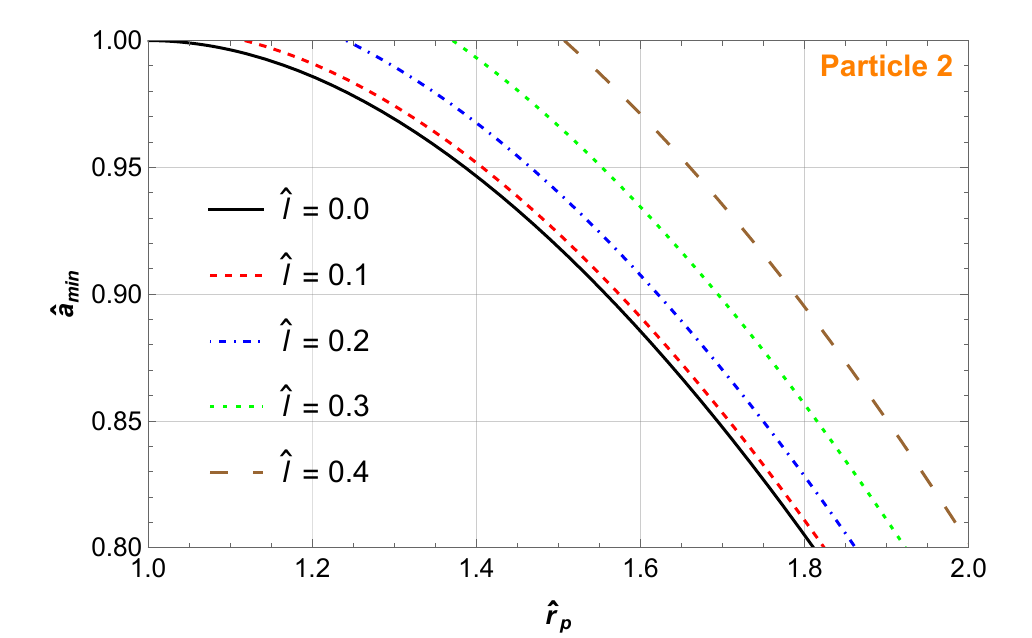}
    \caption{The plot demonstrates the minimum spin lower limits as a function of the decay radius $\hat{r}_p$ for the different values of the $\hat{l}$ parameter.}
    \label{fig:p1}
\end{figure}

Using the Eq.~(\ref{eff}), one can define the specific energy as follows~\cite{2018igrc.book.....B,Hartle_2021}
\begin{equation}\label{energy}
\hat{\cal{E}}=-\frac{g_{tt}+\Omega g_{t\phi}}{\sqrt{-g_{tt}-2g_{t\phi}\Omega-g_{\phi\phi}\Omega^2}}\,,
\end{equation}
where $\Omega$ refers to the angular velocity of a particle following the co-rotating Keplerian orbit around the Kerr-Taub-NUT BH. It can be written in the following form
\begin{equation}
\Omega=\frac{-g_{t\phi,r}+\sqrt{\{-g_{t\phi,r}\}^2-\{g_{\phi\phi,r}\}\{g_{tt,r}\}}}{g_{\phi\phi,r}}\,.
\end{equation}
Actually, we can directly use the following analytical expression for it, which has already been calculated for the Kerr-Taub-NUT spacetime in~\cite{Chakraborty_2018}
\begin{equation}
\Omega=\frac{\sqrt{M(r^2-l^2)+2l^2r}}{\sqrt{r}(r^2+l^2)+a\sqrt{M(r^2-l^2)+2l^2r}}\,.
\end{equation}
The minimum lower limit of spin for particle 0 can be obtained by solving Eq.~(\ref{energy}) by considering the marginally bound orbit, ${\cal \hat{E}}=1$. For $\hat{E}_0>1$, this limit is governed by particle 2, and is situated at the co-rotating photon sphere radius. As stated in Ref.~\cite{Chakraborty_2019}, the prograde photon sphere radius fulfills the following condition
\begin{equation}
 r^3-3M r^2-3l^2r+Ml^2+2a(Ar)^{\frac{1}{2}}=0
\end{equation}
where $A=M(r^2-l^2)+2l^2r$. The minimum value of the spin for particle 2 can be found by solving the above equation with respect to $a$. Finally, one can define the minimum spin lower limit for particle 1 by using the critical case $\hat{r}_p=r_+$, which is found from the condition that a turning point exists for this particle. In Fig.~\ref{fig:p1}, we demonstrate the dependence of the minimum spin lower limit $\hat{a}_{min}$ of particles on the decay radius $\hat{r}_p$ for the different values of the $l$ parameter. One can see from this figure that the range of the decay radius varies with the ergosphere as a function of the $l$ parameter. Also, the values of the minimum spin lower limit increase with the rise of the $l$ parameter. 
\begin{figure}
    \centering
    \includegraphics[scale=0.33]{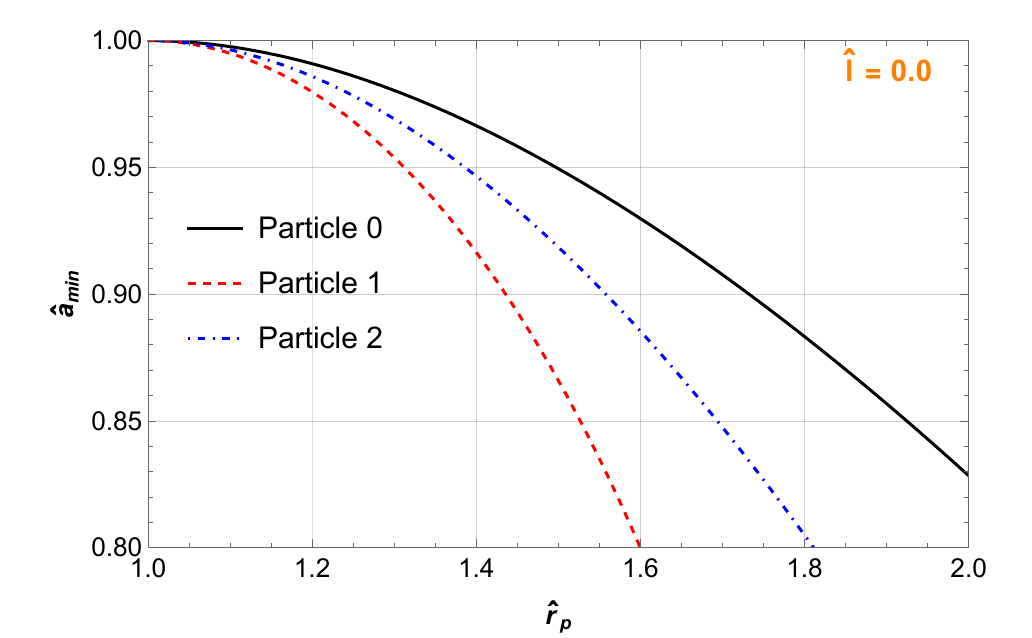}
    \includegraphics[scale=0.33]{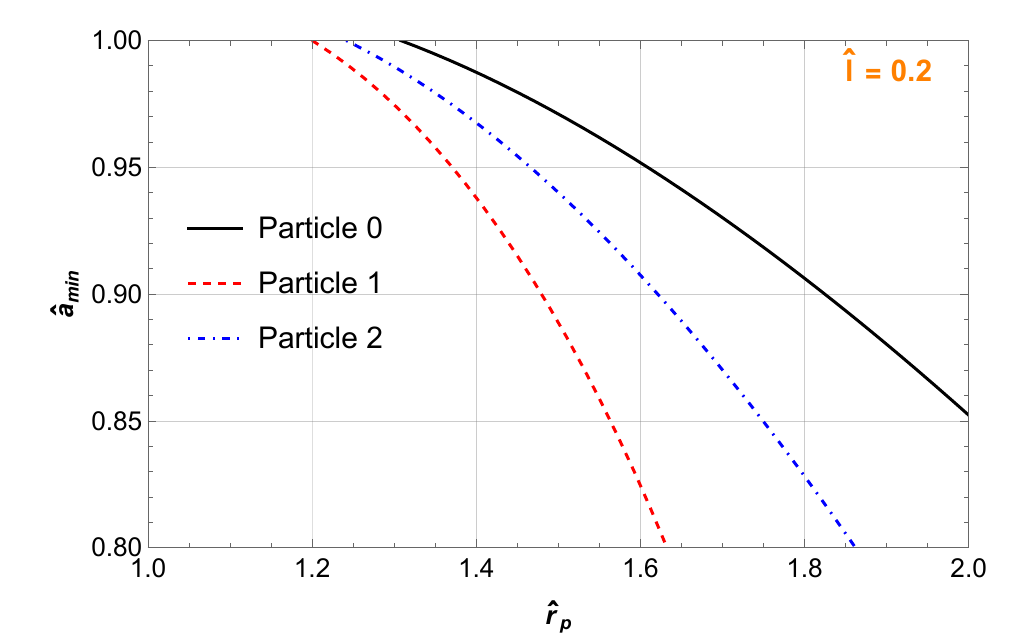}
    \includegraphics[scale=0.33]{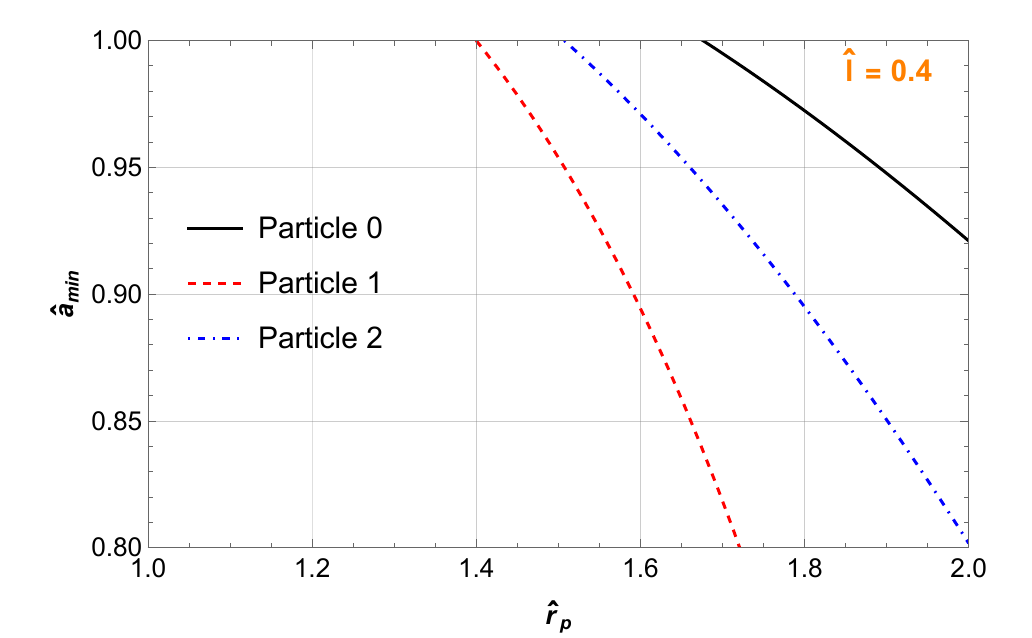}
    \caption{The plot compares the minimum spin lower limits for three particles.}
    \label{fig:p2}
\end{figure}
After that, we show the minimum spin lower limit as a function of the decay radius for each particle in the fixed values of the $l$ parameter in Fig.~\ref{fig:p2}. The results indicate that the values satisfy the following condition
\begin{align}
    \hat{a}_{min,1}<\hat{a}_{min,2}<\hat{a}_{min,0}\,,
\end{align}
and it means that particle 0 governs the spin lower limit for terminating the iteration. Note that this limit is not fixed because the mass $M$ and $l/M$ increase and decrease, respectively, with respect to each iteration.

\section{Numerical results}\label{sec:numerical}

This part is devoted to the numerical results of the repetitive Penrose process. We consider the $\hat{E}_0=1$ case in order to maximize the energy return to investment (see,~\cite{Ruffini:2024dwq}). We also adopt the similar values from Ref.~\cite{Ruffini:2024dwq} as $\hat{p}_{\phi,1}=-19.434$, $\nu=\mu_2/\mu_1=0.78345$ and $\mu_0=10^{-2}M$. 
\begin{table}[t]
    \centering
   \resizebox{1.0\textwidth}{!}{%
    \begin{tabular}{ccccccccccccccc}
    \hline
    \hline
    $n$ & $\frac{M_n}{M_0}$ & $\hat{a}_n$ &$\hat{l}_n$ & $\frac{\mu_{1,n}}{\mu_0}$ & $\frac{E_{1,n}}{\mu_0}$ & $\frac{E_{extractable,n}}{M_0}$ & $\frac{E_{extracted,n}}{M_0}$ & $\frac{M_{irr,n}}{M_0}$ & $\xi_n$ & $\Xi_n$ & $\hat{a}_{min,0,n}$\\
    \hline
        0 & 1.000000 & 1.000000 & 0.200000 & 0.0208248 & -0.0994962 & 0.218975 & 0.000000  & 0.781025  & -- & 0.000000 & 0.987499 \\
 1 & 0.999005 & 0.997938 & 0.200199 & 0.0207192 & -0.0995059 & 0.216472 & 0.000995  &  0.782533 & 0.0994962 & 0.397582 & 0.987540\\
 2 & 0.998010 & 0.995890 & 0.200399 & 0.0206132 & -0.0995157 & 0.214168 & 0.001990  &  0.783842 & 0.0995011 & 0.413967 & 0.987582\\
 3 & 0.997015 & 0.993857 & 0.200599 & 0.0205068 & -0.0995255 & 0.212036 & 0.002985  & 0.784979  &  0.0995060 & 0.430195 & 0.987624\\
 4 & 0.996020 & 0.991839 & 0.200799 & 0.0203999 & -0.0995352 & 0.210057 & 0.003980  & 0.785963 & 0.0995108 & 0.446316 & 0.987666\\
 5 & 0.995024 & 0.989836 & 0.201000 & 0.0202925 & -0.0995450 & 0.208214 & 0.004976  & 0.786810 & 0.0995157 & 0.462373 & 0.987708\\
 6 & 0.994029 & 0.987848 & 0.201201 & 0.0201846 & -0.0995548 & 0.206493 & 0.005971  & 0.787536  & 0.0995206 & 0.478402 & 0.987750\\
 \hline
 \textcolor{red}{7} & \textcolor{red}{0.993033} & \textcolor{red}{0.985876} & \textcolor{red}{0.201403} & \textcolor{red}{0.0200760} & \textcolor{red}{-0.0995646} & \textcolor{red}{0.204885} & \textcolor{red}{0.006967}  & \textcolor{red}{0.788148} & \textcolor{red}{0.0995255} & \textcolor{red}{0.494436} & \textcolor{red}{0.987792}\\
 \hline
 \hline
    \end{tabular}
    }
    \caption{The table shows the change of the parameters for the repetitive Penrose process. Here, we set initially $\hat{l}=0.2$ and $\hat{r}_p=1.4$.}
    \label{tab}
\end{table}
\begin{table}[t]
    \centering
   \resizebox{1.0\textwidth}{!}{%
    \begin{tabular}{ccccccccccccccc}
    \hline
    \hline
    $n$ & $\frac{M_n}{M_0}$ & $\hat{a}_n$ &$\hat{l}_n$ & $\frac{\mu_{1,n}}{\mu_0}$ & $\frac{E_{1,n}}{\mu_0}$ & $\frac{E_{extractable,n}}{M_0}$ & $\frac{E_{extracted,n}}{M_0}$ & $\frac{M_{irr,n}}{M_0}$ & $\xi_n$ & $\Xi_n$ & $\hat{a}_{min,0,n}$\\
    \hline
      0 & 1.000000 & 1.000000 & 0.200000 & 0.0220644 & -0.0610558 & 0.218975 & 0.00000000 & 0.781025 & -- & 0.000000 & 0.951846 \\
      
1 & 0.999389 & 0.996929 & 0.200122 & 0.0219517 & -0.0610643 & 0.214693 & 0.00061056 & 0.784697 & 0.0610558 & 0.142572 & 0.951873 \\

2 & 0.998779 & 0.993874 & 0.200245 & 0.0218391 & -0.0610728 & 0.210786 & 0.00122120 & 0.787993 & 0.0610600 & 0.149126 & 0.951899 \\

3 & 0.998168 & 0.990836 & 0.200367 & 0.0217267 & -0.0610813 & 0.207190 & 0.00183193 & 0.790978 & 0.0610643 & 0.155446 & 0.951926 \\

4 & 0.997557 & 0.987815 & 0.200490 & 0.0216144 & -0.0610898 & 0.203857 & 0.00244274 & 0.793701 & 0.0610685 & 0.161574 & 0.951953 \\

5 & 0.996946 & 0.984810 & 0.200613 & 0.0215023 & -0.0610982 & 0.200749 & 0.00305364 & 0.796197 & 0.0610728 & 0.167543 & 0.95198 \\

6 & 0.996335 & 0.981821 & 0.200736 & 0.0213902 & -0.0611066 & 0.197838 & 0.00366462 & 0.798497 & 0.0610770 & 0.173378 & 0.952007 \\

7 & 0.995724 & 0.978849 & 0.200859 & 0.0212783 & -0.061115 & 0.195102 & 0.00427569 & 0.800623 & 0.0610813 & 0.179100 & 0.952034 \\

8 & 0.995113 & 0.975894 & 0.200982 & 0.0211666 & -0.0611233 & 0.192520 & 0.00488684 & 0.802593 & 0.0610855 & 0.184725 & 0.952061 \\

9 & 0.994502 & 0.972955 & 0.201106 & 0.0210549 & -0.0611317 & 0.190078 & 0.00549807 & 0.804424 & 0.0610897 & 0.190267 & 0.952088 \\

10 & 0.993891 & 0.970033 & 0.201229 & 0.0209433 & -0.0611399 & 0.187763 & 0.00610939 & 0.806128 & 0.0610939 & 0.195738 & 0.952115 \\

11 & 0.993279 & 0.967127 & 0.201353 & 0.0208318 & -0.0611482 & 0.185563 & 0.00672079 & 0.807716 & 0.0610981 & 0.201148 & 0.952142 \\

12 & 0.992668 & 0.964238 & 0.201477 & 0.0207204 & -0.0611565 & 0.183468 & 0.00733227 & 0.809199 & 0.0611022 & 0.206505 & 0.952169 \\

13 & 0.992056 & 0.961366 & 0.201601 & 0.0206091 & -0.0611647 & 0.181472 & 0.00794383 & 0.810585 & 0.0611064 & 0.211816 & 0.952196 \\

14 & 0.991445 & 0.958510 & 0.201726 & 0.0204978 & -0.0611729 & 0.179565 & 0.00855548 & 0.811888 & 0.0611106 & 0.217088 & 0.952224 \\

15 & 0.990833 & 0.955671 & 0.201850 & 0.0203866 & -0.0611811 & 0.177742 & 0.00916721 & 0.813091 & 0.0611147 & 0.222327 & 0.952251 \\

16 & 0.990221 & 0.952849 & 0.201975 & 0.0202754 & -0.0611892 & 0.175998 & 0.00977902 & 0.814223 & 0.0611189 & 0.227538 & 0.952278 \\
\hline
\textcolor{red}{17} & \textcolor{red}{0.989609} & \textcolor{red}{0.950044} & \textcolor{red}{0.2021} & \textcolor{red}{0.0201643} & \textcolor{red}{-0.0611973} & \textcolor{red}{0.174326} & \textcolor{red}{0.0103909} & \textcolor{red}{0.815283} & \textcolor{red}{0.061123} & \textcolor{red}{0.232726} & \textcolor{red}{0.952306} \\
 \hline
 \hline
    \end{tabular}
    }
    \caption{The same with the Table~\ref{tab} but for $\hat{r}_p=1.6$.}
    \label{tab2}
\end{table}

Firstly, we consider that the decay radius is initially located at $\hat{r}_p=1.4$, and the initial value for the dimensionless gravitomagnetic charge is $\hat{l}_p=0.2$. The Table~\ref{tab} demonstrates the numerical values of the BH's parameters and the aforementioned quantities of the repetitive Penrose process at each step. As we discussed earlier, the minimum spin lower limit is controlled by particle 0. Therefore, we present the values of it in the last column of the table. The results indicate that there is a slight increase in the values of $\hat{a}_{min,0}$ with every iteration. One can see by comparing the third and last columns of this table that the iteration has stopped at $n=6$. From the 7th iteration onward, the results do not meet the iteration condition. In other words, the values of the $\hat{a}_{min}$ exceed the values of $\hat{a}$. For instance, $\hat{a}_7=0.985876$ and $\hat{a}_{min,0,7}=0.987792$ for the 7th iteration. Moreover, the results in this table fulfill the iteration conditions stated at the beginning of the above section. If we consider the mass deficit condition, we take $\tilde{\mu}_1<1/(1+\mu_2/\mu_1)=0.56$. It can also be seen from the table that $\hat{E}_1$ is negative and $E_{extractable,n}$ is positive for every iteration, and that the irreducible mass is consistently non-decreasing. 

The Table~\ref{tab} reveals that one part of the reduction in extractable energy contributes to the extracted energy, while another fraction goes toward increasing the irreducible mass. The energy utilization efficiency indicates that 47.84\% of the variation in the extractable energy is transformed into extracted energy. 

\begin{figure}
    \centering
    \includegraphics[scale=0.5]{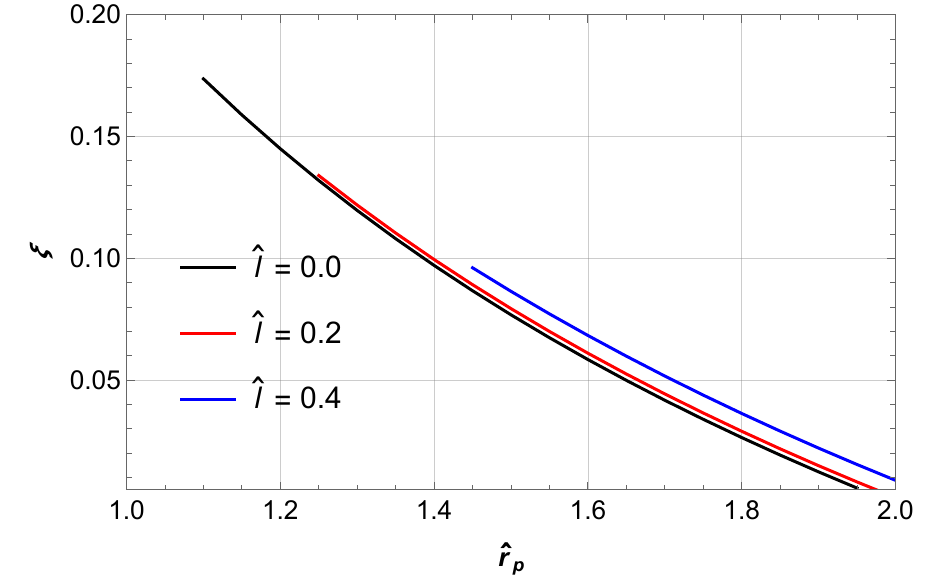}
    \caption{The plot shows the energy return on investment as a function of the decay radius $\hat{r}_p$ for the different values of the $\hat{l}$ parameter after the ending of the repetitive Penrose process.}
    \label{fig:last}
\end{figure}

In addition, we only change the decay radius, and we numerically calculate the similar parameters and quantities for the decay radius located at $\hat{r}_p=1.6$ in Table~\ref{tab2}. As can be seen from this table, all the conditions mentioned above are satisfied here up to the 17th iteration. Because $\hat{a}_{17}=0.950044$ and $\hat{a}_{min,0,17}=0.952306$ for the 17th iteration. It is worth noting that the red color indicates the forced continuation, and these results in Tables~\ref{tab} and~\ref{tab2} are meaningless. According to the results, decreasing the spin of the Kerr-Taub-NUT BH does not allow the extraction of all of its associated rotational energy due to the nonlinear growth of the irreducible mass. It is worth noting that after the repetitive Penrose process ends, the residual extractable energy is $0.206493 M$ and $0.175998 M$ for, respectively, $\hat{r}_p=1.4$ and $\hat{r}_p=1.6$, indicating that a significant amount of energy remains to be extracted through other mechanisms. To be more informative, we plot the energy return on investment as a function of the decay radius for the different values of the $\hat{l}$ parameter.

\section{Conclusions}\label{summary}

In this work, we study the energy extraction from the Kerr-Taub-NUT BH through the repetitive Penrose process. Actually, the maximum spin of the Kerr-Taub-NUT BH $a/M$ exceeds unity due to the gravitomagnetic charge. However, in order to compare the results with those in the Kerr BH case, we set the initial value of the spin as $a/M=1$. First of all, we briefly review the spacetime of the Kerr-Taub-NUT BH, including the horizon and ergoregion structures. Subsequently, we define the extractable energy of the BH using the equation for the irreducible mass. It was found that the values of the extractable energy decrease with the increase of the gravitomagnetic charge. Furthermore, we turn to the theoretical framework of the repetitive Penrose process and the corresponding conditions. For instance, we plot the minimum spin lower limits for the three particles as a function of the decay radius for the different values of the $\hat{l}_p$ parameter. By analyzing them, we find that the spin lower limit is governed by particle 0. After that, we choose initial values for the gravitomagnetic charge and the decay radius as $\hat{l}=0.2$ and $\hat{r}_p=1.4$, and the change in the parameters of the Kerr-Taub-NUT BH due to the repetitive Penrose process and corresponding quantities of this energy extraction mechanism were calculated numerically. Additionally, we perform the same numerical calculations for another decay radius at $\hat{r}_p=1.6$. According to the results, reducing the spin of the BH does not allow all of the corresponding rotational energy to be extracted; the limitation arises from the nonlinear rise of the irreducible mass. To provide more information, we plot the energy return on investment as a function of the $\hat{r}_p$ for the different values of the $\hat{l}$. It was found that the maximum energy return on investment increases as the initial $\hat{l}$ decreases.

\section*{ACKNOWLEDGEMENT}

This research was funded by the National Natural Science Foundation of China (NSFC) under Grant No. U2541210.

\bibliographystyle{apsrev4-1}  

\bibliography{ref}

\end{document}